\documentclass[12pt]{article}
\usepackage{epsfig}
\usepackage{amsfonts}
\begin{document}
\begin{titlepage}
\begin{center}

{\Large Superstatistics: Theory and Applications}

\vspace{2cm}

{\bf C. Beck}

\vspace{0.5cm}

School of Mathematical Sciences, Queen Mary, University of London,
Mile End Road, London E1 4NS, UK

\vspace{3cm}

\end{center}

\abstract{Superstatistics is a superposition of two different statistics
relevant for driven nonequilibrium systems with a stationary state
and intensive parameter fluctuations. It contains Tsallis
statistics as a special case. After briefly summarizing some of the
theoretical aspects, we describe recent applications of this
concept to three different physical problems, namely
a) fully developed hydrodynamic turbulence b) pattern formation in
thermal convection states and c) the statistics of cosmic rays.}

\vspace{1.3cm}

\end{titlepage}

\section{Introduction}

The formalism of nonextensive statistical mechanics
\cite{tsa1,tsa2,tsa3,abe} can be regarded as an embedding of
ordinary statistical mechanics into a more general framework. The
basic idea is that some systems to which ordinary statistical
mechanics is not applicable may still obey a similar formalism if
the Shannon entropy is replaced by more general entropy measures.
Good candidates for these more general entropy measures are the
Tsallis entropies, which depend on a real parameter $q$ and which
reduce to the Shannon entropy for $q\to 1$. For general $q$, one
obtains generalizations of the canonical distributions,
represented by $q$-exponentials, often called Tsallis
distributions. There can be a variety of reasons why ordinary
statistical mechanics is not applicable to a particular system:
There may be long-range interactions, metastability, driving
forces that keep the system out of equilibrium, etc.

Tsallis distributions are indeed observed in a large variety of
physical systems \cite{tsa3,abe,BLS,pla,daniels3,bediaga,cos1}, many
of them being in a driven stationary state far from equilibrium.
An important question is why these or similar distributions are
that often observed in experiments. Can we give dynamical reasons
for the occurence of Tsallis statistics in suitable classes of
nonequilibrium systems?

This is indeed possible. One can easily construct classes of
stochastic differential equations with fluctuating parameters
where one can rigorously prove that they generate Tsallis
statistics \cite{wilk,prl}. Indeed, for many systems the reason
why Tsallis distributions are observed can be easily related to
the fact that there are spatio-temporal fluctuations of an
intensive parameter (e.g.\ the inverse temperature, or a friction
constant, or the amplitude of the Gaussian white noise, or the
energy dissipation in turbulent flows). If these fluctuations
evolve on a long time scale and are distributed according to a
particular distribution, the $\chi^2$-distribution, one ends up
with Tsallis statistics in a natural way. For other distributions
of the intensive parameter, one ends up with more general
statistics: So-called superstatistics \cite{eddie}, which contain
Tsallis statistics as a special case. Generalized entropies
(analogues of the Tsallis entropies) can be defined for these
superstatistics as well \cite{souza,abearbi}, and generalized versions of
statistical mechanics can be also constructed, at least in
principle. It has been shown that the corresponding generalized
entropies are stable \cite{abestab,souza2}.

In this paper we will briefly review the superstatistics concept,
and then show that the corresponding stationary probability distributions
are not only a theoretical construct
but of practical physical relevance.
We will concentrate onto
three physically relevant
examples of applications: Hydrodynamic fully developed turbulence, defect
motion in convection states, and the statistics of cosmic rays.

\section{What is superstatistics?}

\subsection{The basic idea}

Let us give a short introduction to the `superstatistics' concept
\cite{eddie}.
Consider a driven nonequilibrium
systems with spatio-temporal fluctuations of an intensive
parameter $\beta$. This may be the inverse temperature, or a
chemical potential, or a function of the fluctuating energy
dissipation in the flow (for the turbulence application). Locally,
i.e.\ in spatial regions (cells)
where $\beta$ is approximately constant, the system
is described by ordinary statistical mechanics, i.e.\ ordinary
Boltzmann factors $e^{-\beta E}$, where $E$ is an effective energy
in each cell. In the long-term run, the system is described
by a spatio-temporal average over the fluctuating $\beta$.
In this way one
obtains a superposition of two statistics (that of $\beta$ and
that of $e^{-\beta E}$), hence the name `superstatistics'. One may
define an averaged Boltzmann factor $B(E)$ as
\begin{equation}
B(E) =\int_0^\infty f(\beta) e^{-\beta E}d\beta ,
\end{equation}
where $f(\beta)$ is the probability distribution of $\beta$. For so-called
type-A superstatistics, one normalizes this effective Boltzmann
factor and obtains the stationary long-term probability distribution
\begin{equation}
p(E)=\frac{1}{Z}B(E),
\end{equation}
where
\begin{equation}
Z=\int_0^\infty B(E)dE.
\end{equation}
For type-B superstatistics, one includes the $\beta$-dependent
normalization constant into the averaging process. In this case
\begin{equation}
p(E)=\int_0^\infty f(\beta) \frac{1}{Z(\beta)}e^{-\beta E}d\beta ,
\end{equation}
where $Z(\beta)$ is the normalization constant of $e^{-\beta E}$
for a given $\beta$. Both approaches can be easily mapped into
each other, by defining a new probability density
$\tilde{f}(\beta)=c\cdot f(\beta)/Z(\beta)$,
where $c$ is a normalization constant. It is obvious that
Type-B superstatistics with $f$ is equivalent to type-A
superstatistics with $\tilde f$.

As will be proved later, $p(E)$ becomes a Tsallis distribution if
$f(\beta)$ is chosen as a $\chi^2$-distribution. But at the moment
we keep $f(\beta)$ general. Many relevant concepts of
superstatistics can be formulated for general $f(\beta)$.

\subsection{Dynamical realization}

A simple dynamical realization of a superstatistics can be
constructed by considering stochastic differential equations with
spatio-temporally fluctuating parameters \cite{prl}. Consider a
Langevin equation for a variable $u$
\begin{equation}
\dot{u}= \gamma F(u) +\sigma L(t), \label{1}
\end{equation}
where $L(t)$ is Gaussian white noise, $\gamma >0$ is a friction
constant, $\sigma$ describes the strength of the noise, and
$F(u)=-\frac{\partial}{\partial u} V(u)$ is a drift force. If
$\gamma$ and $\sigma$ are constant then the stationary probability
density of $u$ is proportional to $e^{-\beta V(u)}$, where
$\beta:=\frac{\gamma}{\sigma^2}$ can be identified with the
inverse temperature of ordinary statistical mechanics. Most
generally, however, we may let the parameters $\gamma$ and
$\sigma$ fluctuate so that $\beta=\frac{\gamma}{\sigma^2}$ has
probability density $f(\beta)$. These fluctuations are assumed to
be on a long time scale so that the system can temporarily reach
local equilibrium. In this case one obtains for the conditional
probability $p(u|\beta)$ (i.e. the probability of $u$ given some
value of $\beta$)
\begin{equation}
p(u|\beta)=\frac{1}{Z(\beta)}\exp \left\{ -\beta V(u)\right\},
\end{equation}
for the joint probability $p(u,\beta)$ (i.e. the probability to
observe both a certain value of $u$ and a certain value of
$\beta$)
\begin{equation}
p(u,\beta)=p(u|\beta)f(\beta)
\end{equation}
and for the marginal probability $p(u)$ (i.e.\ the probability to
observe a certain value of $u$ no matter what $\beta$ is)
\begin{equation}
p(u)=\int_0^\infty  p(u|\beta)f(\beta)d\beta . \label{99}
\end{equation}
This marginal distribution is the generalized canonical
distribution of the superstatistics considered. The above
formulation corresponds to type-B superstatistics.

Let us now consider a few examples of possible superstatistics, by
considering different examples of distributions $f(\beta)$.
Note that $\beta$ lives on a positive support, so Gaussian distributions of
$\beta$ are unsuitable.

\subsection{$\chi^2$-superstatistics}
One of the most natural choices
of the probability density of $\beta$ is given by the $\chi^2$-distribution
\begin{equation}
f (\beta) = \frac{1}{\Gamma \left( \frac{n}{2} \right)} \left\{
\frac{n}{2\beta_0}\right\}^{\frac{n}{2}} \beta^{\frac{n}{2}-1}
\exp\left\{-\frac{n\beta}{2\beta_0} \right\} . \label{fluc}
\end{equation}
Note that if one adds Gaussian random variables then the sum is
again a Gaussian random variable. But if one adds Gaussian random
variables squared, one gets a $\chi^2$-distributed random
variable. In this sense the $\chi^2$-distribution (also called
$\Gamma$ distribution) is a typical distribution for positive
random variables that naturally arises in many circumstances. Let
us denote the $n$ independent Gaussian random variables as
$X_i,\;i=1,\ldots ,n$ and assume they have average $0$. Then
\begin{equation}
\beta:=\sum_{i=1}^{n} X_i^2 \label{Gauss}
\end{equation}
has the probability density (\ref{fluc}). The average of the
fluctuating $\beta$ is given by
\begin{equation}
\langle \beta \rangle =n\langle X^2\rangle=\int_0^\infty\beta
f(\beta) d\beta= \beta_0
\end{equation}
and the variance by
\begin{equation}
\langle \beta^2 \rangle -\beta_0^2= \frac{2}{n} \beta_0^2.
\end{equation}
For linear drift forces $F(u)=-u$
one obtains for the conditional probability $p(u|\beta)$
\begin{equation}
p(u|\beta)=\sqrt{\frac{\beta}{2\pi}}\exp \left\{ -\frac{1}{2}\beta
u^2\right\},
\end{equation}
and for the marginal probability $p(u)$ a short calculation yields
\begin{equation}
p(u)=\int_0^\infty p(u|\beta)f(\beta)d\beta =
\frac{\Gamma\left(\frac{n}{2}+\frac{1}{2}\right)}{ \Gamma
\left(\frac{n}{2} \right)} \left( \frac{\beta_0}{\pi n}
\right)^{\frac{1}{2}} \frac{1}{\left( 1+\frac{\beta_0}{n}u^2
\right)^{\frac{n}{2}+\frac{1}{2}}}.
\end{equation}
Thus the stochastic differential equation (\ref{1}) with
$\chi^2$-distributed $\beta=\gamma/\sigma^2$ generates the
generalized canonical distributions of nonextensive statistical
mechanics
\begin{equation}
p(u) \sim \frac{1}{\left(
1+\frac{1}{2}\tilde{\beta}(q-1)u^2\right)^{\frac{1}{q-1}}}
\end{equation}
provided the following identifications are made:
\begin{eqnarray}
\frac{1}{q-1}=\frac{n}{2}+\frac{1}{2} &\Longleftrightarrow&
q=1+\frac{2}{n+1}\label{qnn} \\ \frac{1}{2}(q-1)\tilde{\beta}=
\frac{\beta_0}{n} &\Longleftrightarrow&
\tilde{\beta}=\frac{2}{3-q} \beta_0.
\end{eqnarray}

\subsection{Log-normal superstatistics}

Of particular interest for
models of hydrodynamic turbulence is the log-normal distribution
\begin{equation}
f(\beta) = \frac{1}{\beta s \sqrt{2\pi}}\exp\left\{ \frac{-(\log
\frac{\beta}{m})^2}{2s^2}\right\}
\end{equation}
It yields yet another possible superstatistics (see
\cite{reynolds,my-own,cast,suhy} for related turbulence models);
$m$ and $s$ are parameters. The average $\beta_0$ of the above
log-normal distribution is given by $\beta_0=m\sqrt{w}$ and the
variance by $\sigma^2=m^2w(w-1)$, where $w:= e^{s^2}$. One obtains
for linear drift forces $F(u)=-u$ the superstatistics distribution
\begin{equation}
p(u) = \frac{1}{2\pi s }\int_0^\infty d\beta \; \beta^{-1/2}
\exp\left\{ \frac{-(\log \frac{\beta}{m})^2}{2s^2}\right\}
e^{-\frac{1}{2}\beta u^2}.  \label{10}
\end{equation}
The integral
cannot be evaluated in closed form, but it can be easily
numerically evaluated.

\subsection{F-superstatistics}

Another example is superstatistics based on F-distributions.
The F-distribution is given by
\begin{equation}
f(\beta) =\frac{\Gamma ((v+w)/2)}{\Gamma (v/2) \Gamma (w/2)}
\left( \frac{bv}{w} \right)^{v/2} \frac{\beta^{\frac{v}{2}-1}}{(1+
\frac{vb}{w}\beta)^{(v+w)/2}} .
\end{equation}
Here $w$ and $v$ are positive integers and $b>0$ is a parameter.
We note that for $v=2$ we obtain a Tsallis distribution. However,
this is
a Tsallis distribution in
$\beta$-space, not in $E$-space.

The average of $\beta$ is given by
\begin{equation}
\beta_0=\frac{w}{b(w-2)}
\end{equation}
and the variance by
\begin{equation}
\sigma^2=\frac{2w^2(v+w-2)}{b^2 v (w-2)^2(w-4)} .
\end{equation}

Again, the integral leading to the marginal
distribution $p(u)$ cannot be obtained in closed form,
but is easily numerically
evaluated. Superstatistics based on F-distributions has been
studied in more detail in \cite{sattin}, and possible applications
in plasma physics were sketched there.

\subsection{General properties of superstatistics}

For small $E$, all superstatistics have been shown to have the
same first-order corrections to the Boltzmann factor of ordinary
statistical mechanics as Tsallis statistics has \cite{eddie}. For
moderately large $E$, one often observes similar behaviour as for
Tsallis statistics (see \cite{suhy} for examples). On the other
hand, the extreme tails of $p(E)$ for very large $E$ are very
different for the various superstatistics. Tsallis distributions
decay with a power law for $E\to \infty$, general superstatistics
can have all kinds of asymptotic decays.

For {\em any} superstatistics, not only Tsallis statistics, one
can generally define a parameter $q$ by the relation
\begin{equation}
(q-1)\beta_0^2=\sigma^2,
\end{equation}
or equivalently
\begin{equation}
q=\frac{\langle \beta^2 \rangle}{\langle \beta \rangle^2}.
\label{30}
\end{equation}
For $\chi^2$-superstatistics, this $q$ coincides with the index
$q$ of the Tsallis entropies. For the log-normal distribution we
obtain from eq.~(\ref{30}) $q=w$ and for the F-distribution
$q=1+\frac{2(v+w-2)}{v(w-4)}$. The meaning of this generally
defined parameter $q$ is that $\sqrt{q-1}=\frac{\sigma}{\beta_0}$
is just the coefficient of variation of the distribution
$f(\beta)$, defined by the ratio of standard deviation and mean.
If there are no fluctations of $\beta$ at all, we obtain $q=1$,
i.e.\ ordinary statistical mechanics. Our formula (\ref{30})
relating $q$ and the variance of the $\beta$ fluctuations is valid
for both type-A and type-B superstatistics, just that one has to
form averages with either $f$ and $\tilde{f}$.

As recently shown by Tsallis and Souza \cite{souza, souza2},
general superstatistics maximize more general classes of
entropy-like functions subject to suitable constraints. This can
be used as a starting point to construct a generalized statistical
mechanics for general superstatistics.

\subsection{Many particles}

So far our dynamical realization of a superstatistics in terms of
a stochastic differential equation was written down for one test
particle in one dimension. To generalize to $N$ particles in $d$
space dimensions, we may consider coupled systems of equations
with fluctuating friction forces, as given by
\begin{equation}
\dot{\vec{u}}_i=-\gamma_i \vec{F}_i(\vec{u}_1,\ldots ,
\vec{u}_N)+\sigma_i {\vec L}_i(t) \;\;\;\;\;\;\;i=1,\ldots , N.
\label{general}
\end{equation}
Suppose that a potential $V(\vec{u_1},\ldots ,\vec{u_N})$ exists
for this problem such that $\vec{F}_i=-\frac{\partial}{\partial
\vec{u}_i}V$. We can then proceed to marginal stationary
distributions in a similar way as before. One just has to specify
the statistics of the $\beta_i$ in the vicinity of each particle
$i$.

If all $\beta_i=\frac{\gamma_i}{\sigma^2_i}$ are given by the same
fluctuating random variable $\beta_i=\beta$ the integration yields
marginal distributions of the form
\begin{equation}
p(\vec{u}_1, \ldots , \vec{u}_N)= \int_0^\infty
\frac{f(\beta)}{Z(\beta)} e^{-\beta V(\vec{u}_1,\ldots
,\vec{u}_N)}d\beta \label{dense2}
\end{equation}
(type-B superstatistics). The result depends on both, $f(\beta)$
and $Z(\beta)$.

Let us consider an example. We consider a partition function
$Z(\beta)$ whose $\beta$-dependence is of the form
\begin{equation}
Z(\beta )= \int d\vec{u}_1 \cdots d\vec{u}_Ne^{-\beta V} \sim
\beta^x e^{-\beta y}, 
\end{equation}
where $x$ and $y$ are some suitable numbers. We consider
$\chi^2$-superstatistics for this problem.
The marginal distributions are obtained as
\begin{equation}
p(\vec{u_1},\ldots ,\vec{u_N})\sim \frac{1}{(1+\tilde{\beta}(q-1)
V(\vec{u_1},\ldots ,\vec{u}_N))^\frac{1}{q-1}}\label{dense}
\end{equation}
where
\begin{equation}
q=1+\frac{2}{n-2x} \label{qx}
\end{equation}
and
\begin{equation}
\tilde{\beta}=\frac{\beta_0}{1+(q-1)(x-\beta_0 y)}. \label{betax}
\end{equation}

Is the assumption of a single fluctuating $\beta$ realistic
for many particles? No.
In many physical applications, the various particles will be
dilute and only weakly interacting. Hence in this case $\beta$ is
expected to fluctuate spatially in such a way that the local
inverse temperature $\beta_i$ surrounding one particle $i$ is
almost independent from the local $\beta_j$ surrounding another
particle $j$. Moreover, the potential is approximately just a sum
of single-particle potentials $V(\vec{u}_1,\ldots ,\vec{u}_N)
=\sum_{i=1}^NV_s(\vec{u}_i)$. In this case integration over all
$\beta_i$ leads to marginal densities of the form
\begin{equation}
p(\vec{u}_1, \ldots ,\vec{u}_N)= \int_0^\infty d\beta_1 \cdots
 \int_0^\infty d\beta_N \prod_{i=1}^N \frac{f(\beta_i)}{Z_s(\beta_i)}
e^{-\beta_i V_s(\vec{u}_i)}, \label{dilute2}
\end{equation}
and for $\chi^2$-superstatistics one obtains
\begin{equation}
p(\vec{u}_1,\ldots ,\vec{u}_N)\sim \prod_{i=1}^N
\frac{1}{(1+\tilde{\beta}(q-1) V_s(\vec{u}_i))^\frac{1}{q-1}}.
\label{dilute}
\end{equation}
This means the $N$-particle nonextensive system reduces to products of
1-particle nonextensive systems.

We notice that for many particles there is no unique answer to
what the physically relevant generalized canonical distributions
are --- it depends on the spatio-temporal statistics of the
$\beta_i$. Systems of very high particle density may be better
described by (\ref{dense2}), dilute systems better by
(\ref{dilute2}). Our physical applications in section 3 mainly
relate to the form (\ref{dilute2}).

\subsection{Further generalizations}

We just briefly mention some further generalizations, without
working them out in much detail.

\begin{itemize}
\item We may regard the temperature $T$ rather than $\beta =T^{-1}$ as the
fundamental variable, and can then take for $T$ a
$\chi^2$-distribution, log-normal distribution, $F$-distribution,
etc. This case is easily reduced to the framework studied so far,
since any chosen probability distribution $f_T$ of $T$ implies a
corresponding probability distribution $f_\beta$ of $\beta$ given
by
\begin{equation}
f_\beta (\beta) =f_T (T) | \frac{dT}{d\beta} | =f_T (\beta^{-1})
\frac{1}{\beta^2}.
\end{equation}

\item For each particle there may be several different
intensive parameters that fluctuate, for example not only the inverse
temperature $\beta$ but also the chemical potential $\mu$
may fluctuate in a spatio-temporal way. In order to proceed to
the marginal distribution,
one then simply has to do two integrals (over $\beta$ and $\mu$)
rather than
one.

\item Rather than starting from ordinary Boltzmann factors
$e^{-\beta E}$ and averaging these over a fluctuating $\beta$, we
could also construct a superstatistics where the statistics in the
single cells is already given by Tsallis statistics, i.e. one
would be $\beta$-averaging $q$-exponentials $e_q^{-\beta E}$ in
this case. This is related to the case of two fluctuating
parameters discussed above: We may think of cases where the
integration over the fluctuating $\mu$ yields $q$-exponentials,
which then still have to be averaged over $\beta$.

\end{itemize}

\section{Applications}

\subsection{Fully developed hydrodynamic turbulnce}

In the turbulence application, $u$ stands for a local velocity
{\em difference} in the turbulent flow. On a very small time
scale, this velocity difference is essentially the acceleration.
The basic idea is that turbulent velocity differences locally
relax with a certain damping constant $\gamma$ and are at the same
time driven by rapidly fluctuating chaotic force differences,
which in good approximation are modelled by Gaussian white noise.
One knows that in turbulent flows the energy dissipation
fluctuates in space and time. Moreover, it is known since the
early papers by Kolmogorov in 1962 that the probability density of
energy dissipation is approximately log-normal in a turbulent
flow. Hence, if $\beta$ is a simple power-law function of the
energy dissipation, this implies a lognormally distributed
$\beta$. We thus end up in a most natural way with log-normal
superstatistics.

Fig.~1 shows an experimentally measured $p(u)$ of velocity
differences $u$ at scale $r=92.5\; \eta$ in a turbulent
Taylor-Couette flow, as obtained by Swinney et al.\cite{BLS}.
$\eta$ denotes the Kolmogorov length scale. The data have been
rescaled to variance 1. The dashed line is the prediction of
log-normal superstatistics. Apparently, there is excellent
agreement between the measured density and log-normal
superstatistics as given by eq.~(\ref{10}). The fitting parameter
for this example is $s^2=0.28$. This is the only fitting
parameter, $m=\sqrt{w}$ is fixed by the condition of variance 1 of
$p(u)$.

Fig.~2 shows the measured probability density of the acceleration
of a Lagrangian test particle in a turbulent flow as obtained in
the experiment of Bodenschatz et al. \cite{boden1,boden2}.
Log-normal superstatistics with $s^2=3.0$ yields a good fit. Since
Bodenschatz's data reach rather large accelerations $a$ (in units
of the standard deviation), the measured tails of the
distributions allow for a sensitive distinction between various
superstatistics. Though $\chi^2$-superstatistics ($=$ Tsallis
statistics) yields also a reasonably good fit of the data for
moderately large accelerations, for the extreme tails log-normal
superstatistics seems to do a better job. The main difference
between $\chi^2$-superstatistics and log-normal superstatistics is
the fact that $p(a)$ decays with a power law for the former ones,
whereas it decays with a more complicated logarithmic law for the
latter ones.

\subsection{Chaotic defect motion in inclined layer convection}

Let us now consider another physically relevant example, so-called
`defect turbulence'. Defect turbulence shares with ordinary
turbulence only the name, otherwise it is very different. It is a
phenomenon related to convection and has nothing to do with fully
developed hydrodynamic turbulence. Consider a Raleigh-Benard
convection experiment: A liquid is heated from below and cooled
from above. For large enough temperature differences, interesting
convection patterns start to evolve. An inclined layer convection
experiment \cite{daniels3,daniels2} is a kind of Raleigh-Benard
experiment where the apparatus is tilted by an angle (say 30
degrees), moreover the liquid is confined between two very narrow
plates. For large enough temperature differences, the convection
rolls evolve chaotically. Of particular interest are the defects
in this pattern, i.e.\ points where two convection rolls merge
into one (see Fig.~3). These defects behave very much like
particles. They have a well-defined position and velocity, they
are created and annihilated (in pairs of course), and one can even
formally attribute a `charge' to them: There are positive and
negative defects. For more details, see \cite{daniels3,daniels2}
and references therein.

The probability density of defect velocities has been quite
precisely measured. As shown in Fig.~4, it quite precisely
coincides with a Tsallis distribution with $q \approx 1.46$. Also
other dynamical properties of the defects coincide quite well with
typical predictions of nonextensive models \cite{daniels3}. In
total, these defects seem to behave very much like an ideal gas of
nonextensive statistical mechanics with entropic index $q\approx
1.5$. Defect turbulence thus may serve as an example where
generalized versions of statistical mechanics are not only
mathematically beautiful but physically useful. Note that this is
an application to a physical system far from equilibrium, where
ordinary statistical mechanics has little to say.

Our dynamical realization in terms of the generalized Langevin
equation (\ref{1}) with $F(u)=-u$
and fluctuating effective friction $\gamma$ makes sense as a very
simple model
for the defect velocity $u$.
While ordinary Brownian particles have
constant damping due to Stokes' law $\gamma = \frac{6\pi \nu \rho
a}{m}$, where $\nu$ is the kinematic viscosity of the liquid, $\rho$
is its density, $m$ is the mass of the particle and $a$ is the
radius of the particle, defects are no ordinary particles: they
have neither a well-defined mass $m$ nor a well-defined radius $a$
and thus one expects that there is an ensemble of damping
constants which depend on the topology of the defect and its
fluctuating environment. In particular, the fastest velocities
result from circumstances in which the defect is moving in a local
environment with only a very small effective damping $\gamma$
acting. The driving forces $L(t)$ are hardly damped during such a
time interval, and lead to very large velocities for a limited
amount of time, until another region with another $\gamma$ is
reached.

The experimentally observed value $q\approx 1.46...1.50$ for the defect
statistics means according to eq.~(\ref{qnn}) that there are
effectively about three independent degrees of
freedom that contribute to the fluctuating local defect
environment. We do not know where these three effective degrees of
freedom come from, but a very simple picture would be that the
fluctuating environment of the defect is mainly characterized by
the states of the three convection rolls that merge when forming a
defect.

\subsection{Statistics of cosmic rays}

Our third example is from high energy physics. We will proceed to
extremely high temperatures, where (similar as in defect
turbulence) particles are created and annihilated in pairs.
Nonextensive statistical mechanics has been shown to work well for
reproducing experimentally measured cross sections in $e^+e^-$
collider experiments \cite{bediaga,e+e-}. Here we apply it to high
energy collision processes as induced by astrophysical sources,
leading to the creation of cosmic ray particles \cite{pdg} that
are ultimately observed on the earth. The idea to apply
nonextensive statistics to the measured cosmic ray spectrum was
first presented in \cite{cos1}, some related work can also be
found in \cite{cos2,cos3}.

Experimental data of the measured cosmic
ray spectrum
\cite{chicago} are
shown in Fig.~5. Also shown is a curve that
corresponds to a prediction of nonextensive statistical mechanics.
Up to energies of $10^{16}$ eV, the measured flux rate of cosmic
ray particles with a given energy is well fitted by a generalized
canonical distribution of the form
\begin{equation}
p(E)=C \cdot \frac{E^2}{(1+\tilde{\beta}(q-1)E)^{1/(q-1)}}.
\label{can}
\end{equation}
This is an F-distribution in $E$. $E$ is
the energy of the particles,
\begin{equation}
E=\sqrt{c^2p_x^2+c^2p_y^2+c^2p_z^2+m^2c^4},
\end{equation}
$\tilde{\beta}=\tilde{T}^{-1}$ is an effective inverse temperature
variable, and $C$ is a constant representing the total flux rate.
For relativistic particles the rest mass $m$ can be neglected and
one has $E\approx c |\vec{p}|$. The distribution (\ref{can}) is a
$q$-generalized relativistic Maxwell-Boltzmann distribution in the
formalism of nonextensive statistical mechanics. The Tsallis
distribution is multiplied with $E^2$, taking into account the
available phase space volume. As seen in Fig.~5, the cosmic ray
spectrum is very well fitted by the distribution (\ref{can}) if
the entropic index is chosen as $q=1.215$ and and if the effective
temperature parameter is given by
$\tilde{T}=\tilde{\beta}^{-1}=107$ MeV.

The above effective temperature is of the same order of magnitude
as the so-called Hagedorn temperature $T_H$ \cite{hage}, an effective
temperature well known from collider experiments. The Hagedorn
temperature is much smaller than the center-of-mass energy
$E_{CMS}$ of a typical collision process and represents a kind of
`boiling temperature' of nuclear matter at the confinement phase
transition. It is a kind of maximum temperature that can be
reached in a collision experiment. Even largest $E_{CMS}$ cannot
produce a larger average temperature than $T_H$ due to the fact
that the number of possible particle states grows exponentially.
The Hagedorn theory of scattering processes is known to work well
for energies $E_{CMS}<10$ GeV, whereas for larger energies there
is experimental evidence from various collision experiments that
power-law behaviour of differential cross sections sets in. This
power-law is not contained in the original Hagedorn theory but can
be formally incorporated if one considers a nonextensive extension
of the Hagedorn theory \cite{e+e-}.

Let us now work out
the assumption that the nonextensive behaviour of the
measured cosmic ray spectrum is due to
fluctuations of temperature.
Assume that locally, in the creation
process of some cosmic ray particle, some value of the fluctuating inverse
temperature $\beta$ is given. We then expect the momentum of a
randomly picked particle in this region to be distributed
according to the relativistic Maxwell-Boltzmann distribution
\begin{equation}
p(E|\beta)=\frac{1}{Z(\beta)}E^2 e^{-\beta E}. \label{max}
\end{equation}
Here $p(E|\beta)$ denotes the conditional probability of $E$ given
some value of $\beta$. We neglect the rest mass $m$ so that
$E=c|\vec{p}|$. The normalization constant is given by
\begin{equation}
Z(\beta)=\int_0^\infty E^2 e^{-\beta E} dE=\frac{2}{\beta^3} .
\end{equation}
Now assume that $\beta$ is $\chi^2$-distributed. The
observed cosmic ray distribution at the earth does not contain any
information on the local temperature at which the various
particles were produced. Hence we have to average over all
possible fluctuating temperatures, obtaining the measured energy
spectrum as the marginal distribution
\begin{equation}
p(E)=\int_0^\infty p(E|\beta)f(\beta)d\beta . \label{9}
\end{equation}
The integral (\ref{9}) with $f(\beta)$ given by (\ref{fluc}) and
$p(E|\beta)$ given by (\ref{max}) is easily evaluated and one
obtains eq.~(\ref{can})
with
\begin{equation}
q=1+\frac{2}{n+6} \label{qwert}
\end{equation}
and
\begin{equation}
\tilde{\beta}=\frac{\beta_0}{4-3q}.
\end{equation}
This just corresponds to eqs.~(\ref{qx}) and
(\ref{betax}) for $x=-3$ and $y=0$.

The variables $X_i$ in eq.~(\ref{Gauss}) describe the independent
degrees of freedom contributing to the fluctuating temperature. At
very large center of mass energies, due to the uncertainty
relation $r\sim 1/E_{CMS}$, the probed volume $r^3$ is very small,
and all relevant degrees of freedom in this small volume are
basically represented by the 3 spatial dimensions into which heat
can flow. We may physically interpret $X_i^2$ as the heat loss in
the spatial $i$-direction, $i=x,y,z$, during the collision process
that generates the cosmic ray particle. The more heat is lost, the
smaller is the local $T$, i.e. the larger is the local $\beta$
given by eq.~(\ref{Gauss}). The 3 spatial degrees of freedom yield
$n=3$ or, according to eq.~(\ref{qwert}),
\begin{equation}
q=\frac{11}{9}=1.222. \label{qmax}
\end{equation}
For cosmic rays $E_{CMS}$ is very large, hence we expect a
$q$-value that is close to this asymptotic value. The fit in
Fig.~5 in fact uses $q=1.215$, which agrees with the predicted
value in eq.~(\ref{qmax}) to about 3 digits. It also coincides
well with the fitting value $q=1.225$ used by Tsallis et al.
\cite{cos1} using multi-parameter generalizations of nonextensive
canonical distributions.

For smaller center of mass energies, the volume $r^3$ probed will
be bigger and more effective degrees of freedom within this bigger
interaction region can contribute to the fluctuating temperature.
Hence we expect that for decreasing $E_{CMS}$ $n$ will increase
and $q$ will become smaller than $11/9$. Experimental
data measured in $e^+e^-$ annihilation experiments are in good
agreement with the following parametrization \cite{e+e-}:
\begin{equation}
q(E_{CMS})= \frac{11-e^{-E_{CMS}/E_0}}{9+e^{-E_{CMS}/E_0}}
\end{equation}
with $E_0\approx 45.6$ GeV. Solving for $E_{CMS}$ we get
\begin{equation}
E_{CMS}=-E_0 \log \frac{11-9q}{1+q},
\end{equation}
and putting in the fitted value $q=1.215$ of the cosmic ray
spectrum we can give an estimate of the average center of mass
energy of the ensemble of astrophysical accelerators:
\begin{equation}
E_{CMS}\approx 161\; \mbox{GeV}.
\end{equation}
It should be clear that this is a very rough estimate only,
moreover, it is an average over the unknown ensemble of all
accelerating astrophysical
sources. Some astrophysical objects can definitely accelerate
to larger energies.

The `knee' and `ankle' in the cosmic ray spectrum, occuring at
$E\approx 10^{16}$ eV and $E\approx 10^{19}$ eV, are produced by
other effects which can be easily embedded into this nonextensive
model (see \cite{cos3} for details). The advantage of our 
superstatistics model is
that a concrete prediction for $q$ can be given.

\newpage

\epsfig{file=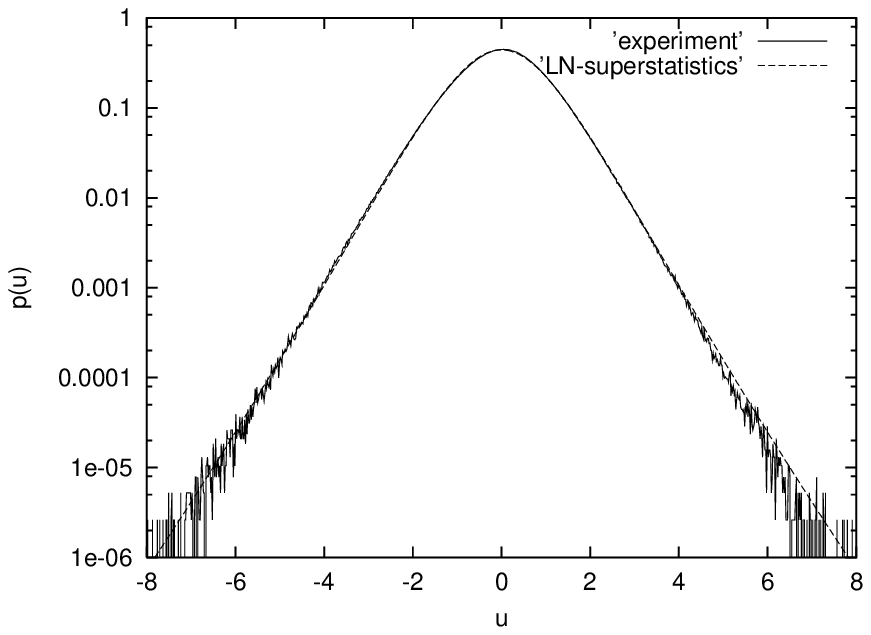}

{\bf Fig.~1} Histogram of velocity differences $u$ as measured
in Swinney's experiment and the
log-normal superstatistics prediction eq.~(\ref{10}) with $s^2=0.28$.

\vspace{1cm}

\epsfig{file=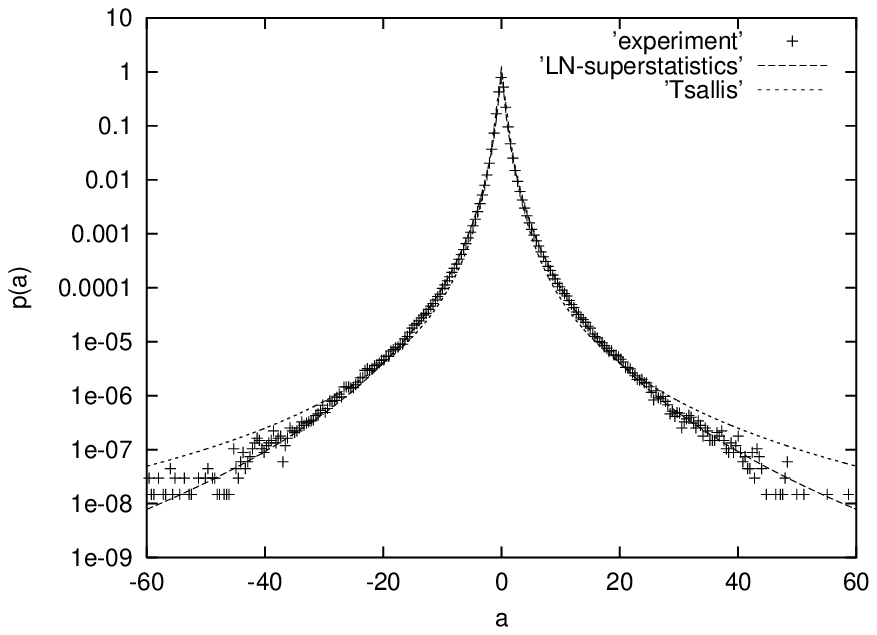}

{\bf Fig.~2} Histogram of accelerations $a$ as measured
in Bodenschatz's experiment,
the log-normal superstatistics prediction
eq.~(\ref{10}) with $s^2=3.0$, and Tsallis statistics with $q=1.5$.

\vspace{1cm}

\epsfig{file=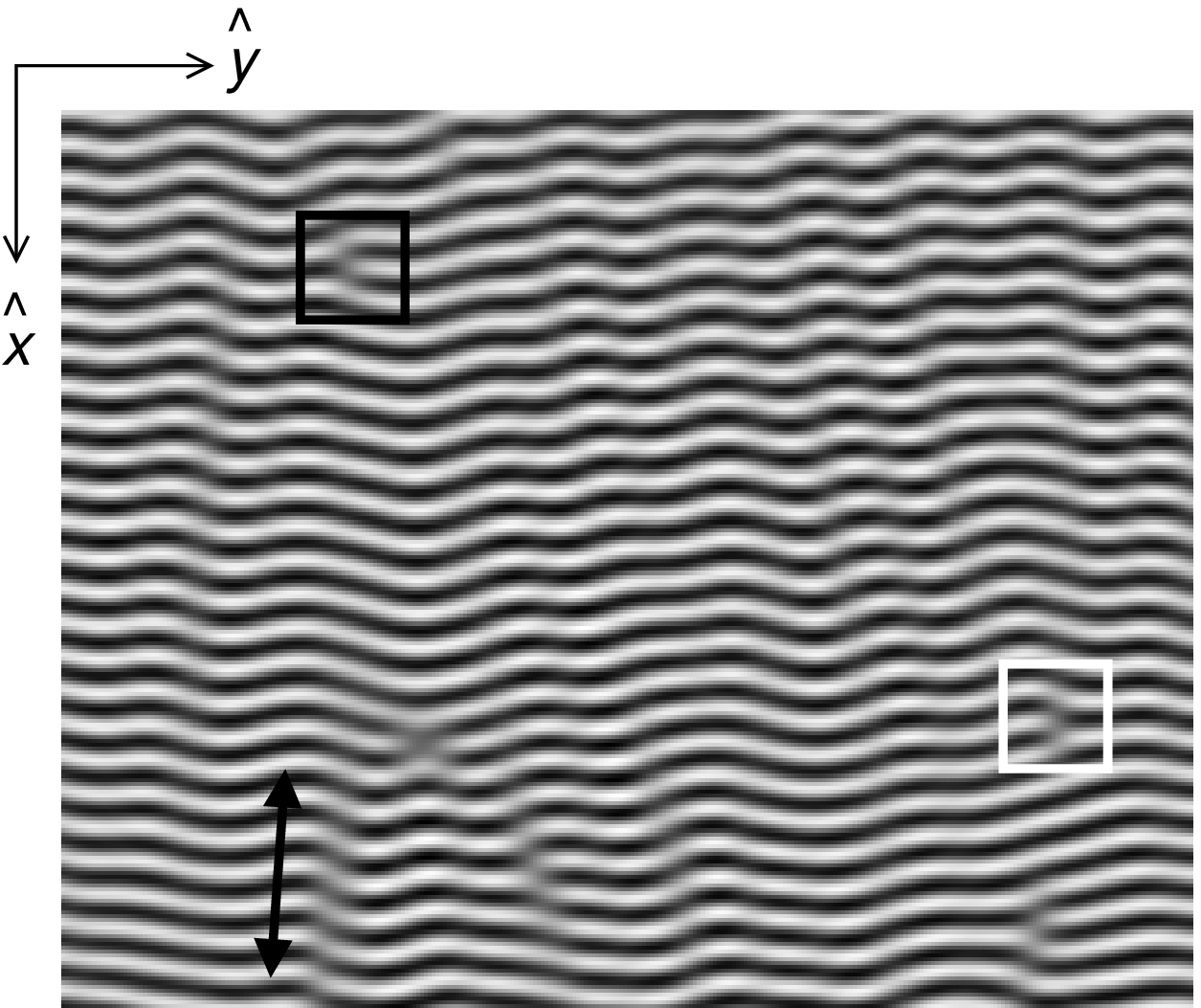}

{\bf Fig.~3} A positive defect (black box) and a negative defect (white box)
as seen in the experiment of Daniels et al. \cite{daniels2}.

\newpage

\epsfig{file=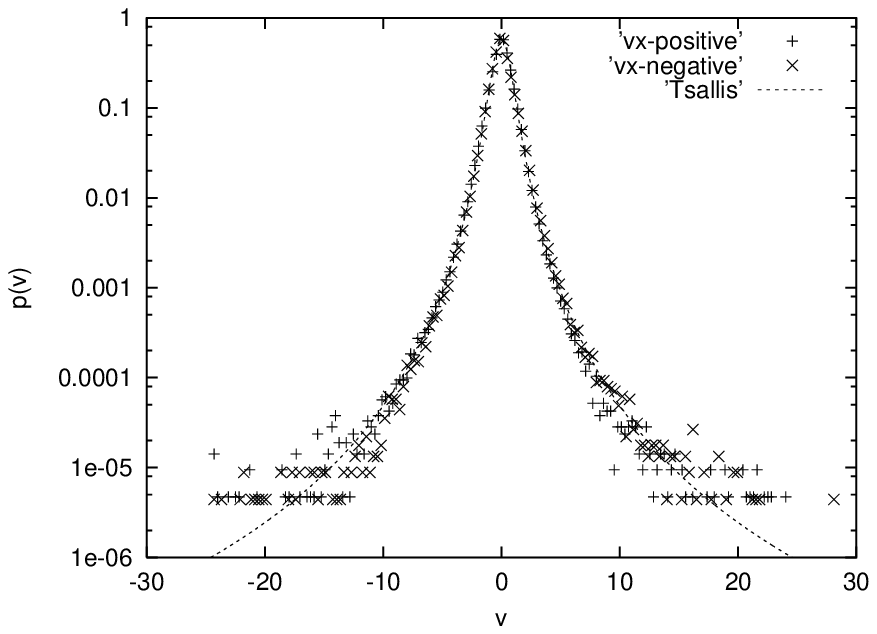}

{\bf Fig.~4} Measured distribution of positive and negative defect
velocities $v_x$ and comparison
with a Tsallis distribution with $q=1.46$. 
The dimensionless temperature difference is
$\frac{\Delta T}{T_c}-1=0.17$. All distributions have been rescaled to variance 1.

\vspace{1cm}

\epsfig{file=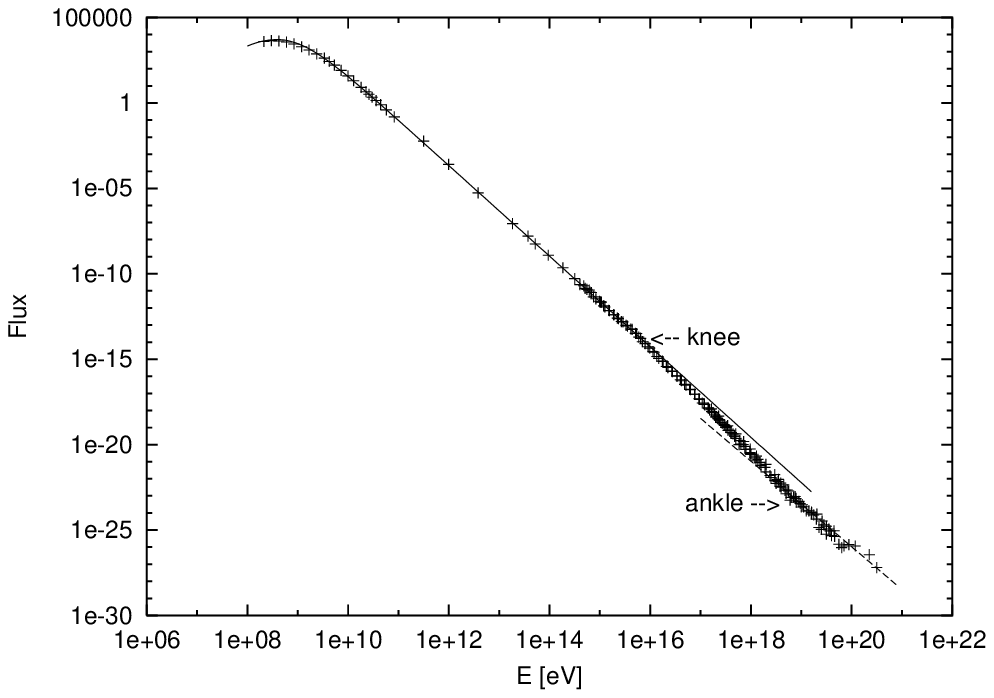}

{\bf Fig.~5} Measured distribution of primary cosmic ray particles
with a given energy $E$ and comparison with the distribution
(\ref{can}) with $q=1.215$ and $\tilde{\beta}^{-1}=107$ MeV
(solid line).

\end{document}